\documentclass[aps,preprint,showpacs]{revtex4}
\usepackage{graphicx,epsfig}
\usepackage{amssymb}
\begin{document}

\title{Geodesics of Hayward black hole surrounded by quintessence}

\author{Omar Pedraza$^1$}
\email{omarp@uaeh.edu.mx}
\author{L. A. L\'opez$^1$}
\email{lalopez@uaeh.edu.mx}
\author{R. Arceo$^2$}
\email{roberto.arceo@unach.mx}
\author{I. Cabrera-Munguia$^{3}$}
\email{icabreramunguia@gmail.com}

\affiliation{$^1$ \'Area Acad\'emica de Matem\'aticas y F\'isica, UAEH, 
Carretera Pachuca-Tulancingo Km. 4.5, C P. 42184, Mineral de la Reforma, Hidalgo, M\'exico.}

\affiliation{$^{2}$ Facultad de Ciencias en F\'isica y Matem\'aticas, Universidad Aut\'onoma de Chiapas, C. P. 29050, Tuxtla Gut\'ierrez, Chiapas, M\'exico.}

\affiliation{$^{3}$ Departamento de F{\'i}sica y Matem\'aticas, Universidad Aut\'onoma de Ciudad Ju{\'a}rez, 32310 Ciudad Ju\'arez, Chihuahua, M\'exico}

\begin{abstract}

Basing on the ideas used by Kiselev, we study the Hayward black hole surrounded by quintessence. By setting for the quintessence state parameter at the special case of $\omega=-\frac{2}{3}$, using the metric of the black hole surrounded by quintessence and the definition of the effective potential, we analyzed in detail the null geodesics for different energies. We also described the horizons of the Hayward black hole surrounded by quintessence as well as the shadow of the black hole.
\end{abstract}


\maketitle

\section{Introduction}

The existence of a central singularity inside a black hole is the difficulty in general relativity because the point singularities will be divergent then the physics laws are incompleteness. Although in the black holes, the singularities in their interior are covered by an event horizon, the prediction made by Hawking several years ago, affirms that the black holes radiation flux causes they shrink until reach the singularity \cite{Giddings:2001bu}. 

To avoid the singularity problem the construction of regular solutions have been proposed, for example, the theory of general relativity coupled to non-linear electrodynamics is one candidate, because, in this theory,  solutions which do not contain a singularity at the center has been constructed, for example  \cite{Bardeen1} \cite{Bronnikov:2000vy}\cite{Dymnikova:2004zc}.

Another idea to generate regular solutions is to consider that a regular solution will contain critical scale, mass, and charge parameters restricted by some value, which depends only on the type of the curvature invariant, this assumption, is called the limiting curvature conjecture \cite{Polchinski:1989ae}.

Following the idea of the limiting curvature Hayward \cite{Hayward:2005gi} proposed a static spherically symmetric black hole that near the origin behaves like a de Sitter spacetime, its curvature invariants being everywhere finite and satisfying the weak energy condition. Variations have been studied as the rotating Hayward \cite{Amir:2015pja} and Hayward charged \cite{Frolov:2016pav}.  Also, diverse investigations have been focused on the properties of the Hayward black hole. For example, in \cite{Perez-Roman:2018hfy}, the interior of the regular Hayward black hole was explored with The Painlev'e-Gullstrand coordinates and in \cite{doi:10.1139/cjp-2019-0572} \cite{Lin:2013ofa} the quasinormal modes were studied.

As the black holes are accepted as part of our universe, the study of consequences of the black holes coexistence with other types of matter or energy is significant. For example, our universe is dominated by dark energy that contains about 70\% of the universe and is responsible for the accelerated expansion of our universe so that the black holes surrounded by dark energy are of interest to researchers. There are alternative models as candidates for dark energy, most of them are based on a scalar field as are quintessence \cite{Capozziello_2006}\cite{PhysRevLett.81.3067}, phantom \cite{PhysRevLett.91.211301}, K-essence \cite{PhysRevLett.85.4438} among others.

Kiselev (2003) \cite{Kiselev:2002dx} present a new static spherically symmetric exact solutions of the Einstein equations for quintessential matter surrounding a black hole, and there have been different investigations applying the Kiselev model. For example, in \cite{Chen_2005}, the quasinormal modes of the Schwarzschild black hole surrounded by the quintessence are studied. Also, the null geodesics for Schwarzschild  surrounded by quintessence in \cite{Fernando:2012ue} is addressed as well as Reissner-Nordstr\"{o}m surrounded by quintessence in \cite{Malakolkalami:2015tsa} and in \cite{Rodrigue:2018lzp} investigate the thermodynamics of Hayward black hole surrounded by quintessence.

In \cite{PhysRevD.101.024022} has investigated that the anisotropic stress-energy leading to Kiselev black hole solution can be represented by being split into a perfect fluid component plus either an electromagnetic component or a scalar field component, then Kiselev black hole fails to represent a perfect fluid spacetime.

For the mentioned above, in the present paper proposed, we study the Hayward black hole surrounded by quintessence matter using the solution obtained by Kiselev, and we analyze the null geodesics for this new solution black hole.
  
The paper is organized as follows: Sec. II shows the Hayward black hole surrounded by quintessence, and some properties are shows, then in section III, we analyze the null geodesics of Hayward surrounded by quintessence. In IV, the examples of null geodesic with different energy are given. In section V, we give a brief analysis of the Hayward black hole's shadow area surrounded by quintessence. Finally, conclusions are given in the last section.
\section{Hayward black hole surrounded by quintessence}

Using the idea of the limiting curvature condition and minimal model, Hayward \cite{Hayward:2005gi} proposed a the regular static spherically symmetric space-time that describes the formation of a  black hole, considering that the Einstein tensor  $G_{\mu\nu}$ has the cosmological constant ($\Lambda$) from  $G \sim - \Lambda g$ as $r \rightarrow 0$ and $\Lambda =3/\epsilon^{2}$ where $\epsilon$ (Hubble length) is a convenient encoding of the central energy density, the effect of $\epsilon$ is that a repulsive force (repulsive core) prevents the singularity. A consequence of including the repulsive core is that the strong energy condition might be violated see \cite{Perez-Roman:2018hfy}. 

The line element of Hayward BH is given by;
\begin{equation}\label{mfa}
ds^2=-f(r)dt^2+\frac{dr^2}{f(r)}+r^2d\theta^2+r^2\sin^2\theta d\phi^2\,,
\end{equation}
where
\begin{equation}
f(r)=1-\frac{2Mr^2}{r^3+2M\epsilon^2}\,.
\end{equation}

Here, $M$ is the parameter of mass, and $\epsilon$ is a parameter related to the cosmological constant. In the limit, $\epsilon\to0$, the metric (\ref{mfa}) reduces to the Schwarzschild black hole. Exist the critical mass $M_{*}=(3\sqrt{3}/4)\epsilon$ and a radius $r_{*}=\sqrt{3}\epsilon$ such that Hayward BH has not horizon if $M<M_{*}$, in the case of $M_{*}=M$ has one horizon at  $r=r_{*}$ and if $M_{*}<M$ has two horizons $r=r_{\pm}$.

 In the investigations of Kiselev \cite{Kiselev:2002dx} about of quintessence and Black holes, Kiselev proposed a new solution of the metric for static and spherically symmetric space-time, considering that the energy-momentum tensor for quintessence should satisfy \cite{Cvetic:2016bxi};

\begin{equation}
T_{\phi}^{ \phi}=T_{\theta}^{ \theta}=-\frac{1}{2}(3\omega +1)T_{r}^{ r}=\frac{1}{2}(3\omega +1)T_{t}^{t}
\end{equation}

where $\omega$ is taken to be a constant and  the dominant energy condition requires $\rho=T_{tt}\geq 0$ ( $\rho$ is the energy density) and $\mid 3\omega + 1 \mid \leq 2$.  Following these ideas, the expression of the metric function of such a black hole surrounded by quintessence is obtained by adding the therm $-c/r^{3\omega+1}$ to the metric of the black holes (see \cite{Fernando:2012ue} \cite{Saleh:2018hba} \cite{Ghaderi:2017wvl} for example). Thus, we can write the function metric for Hayward BH with quintessence as;

\begin{equation}\label{ec.rfc}
f_{\omega}(r)=1-\frac{2Mr^2}{r^3+2M\epsilon^2}-\frac{c}{r^{3\omega+1}}\,,
\end{equation}

where $c$ is a normalization factor and $\omega$ has the range $-1<\omega<-1/3$.  For the Schwarzschild black hole surrounded by quintessence $\epsilon=0$.

\begin{figure}[h]
\begin{center}
\includegraphics [width =0.5 \textwidth ]{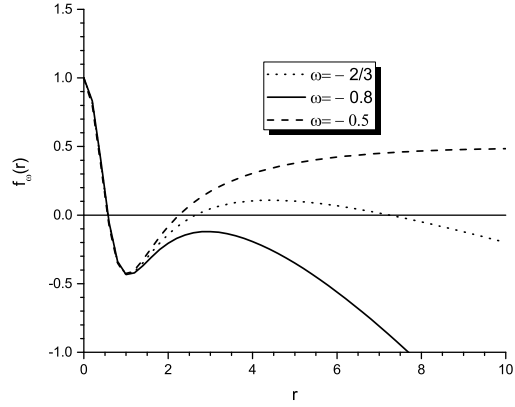}
\end{center}
\caption{Behavior of the function metric $f_{\omega}(r)$ with $M_{*}<M$, $M=1$, $\epsilon =0.5$ and $c=0.1$} \label{Fig1}
\end{figure}

The horizons of Hayward BH with quintessence (Hayward BH-$\omega$) are determined by the roots of the equation $f_{\omega}(r)=0$ In the figure (\ref{Fig1}) shows that in the case of considering of $M_{*}<M$, the Hayward BH-$\omega$ has two horizons for $\omega =-0.5$, and one horizon in the case of $\omega=-0.8$.  For the state of $\omega =-2/3$, $f_{\omega}(r)$ has three roots. Then it is clear how the factor $\omega$ modifies the behavior of the horizons of Hayward BH.

In this paper, we pick $\omega=-\frac{2}{3}$. We can obtain the next relation between the Hayward BH-$\omega$ mass $M_{\omega}$ and its horizon radius $f_{\omega}(r_{h})=0$,

\begin{equation}\label{eq.bqedr1}
M_{\omega}(r_{h})=\frac{1}{2}\frac{r_{h}^3\left(1-cr_{h}\right)}{r_{h}^2+c\epsilon^2r_{h}-\epsilon^2}\,.
\end{equation}

\begin{figure}[h]
\begin{center}
\includegraphics [width =0.7 \textwidth ]{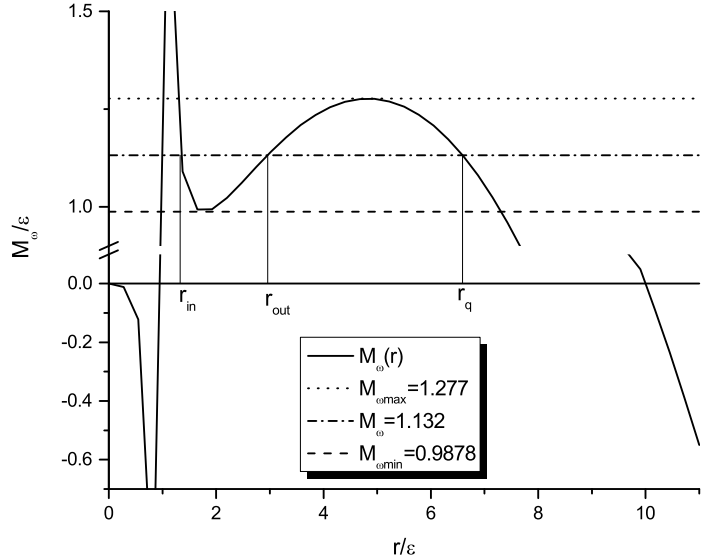}
\end{center}
\caption{The figure shows the horizons of the Hayward BH-$\omega$ with $c\epsilon=0.1$.} \label{Fig2}
\end{figure}

Note that the function $M_{\omega}(r_{h})\to0$ for $r_{h}\to0$, when $r_{h}\to\infty$, $M_{\omega}(r_{h})\to-\infty$ and $M_{\omega}(r_{h})$ is discontinuous at $r_{h}=[\sqrt{c^2\epsilon^2+4}-c\epsilon]\epsilon/2$. The behavior of the function $M_{\omega}(r_{h})$ for general values of $c$ and $\epsilon$ is shows in Fig. (\ref{Fig2}).

 As can be seen from Fig. (\ref{Fig2}), there are two critical value $M_{\omega min}$ and $M_{\omega max}$ for the mass of the Hayward BH-$\omega$ that leads to different physical scenarios: first, for values of $M _{\omega}< M_{\omega min}$ we have a naked singularity. For $M_{\omega}=M_{\omega min}$ ($M_{\omega}=M_{\omega max}$) we have a Hayward BH-$\omega$ with two horizons and for values of $M_{\omega min}<M_{\omega}<M_{\omega max}$ with three horizons. For a value of $M_{\omega}=1.132$, the Hayward BH-$\omega$ has three horizons: the inner horizon $r_{in}$, the event horizon $r_{out}$ ($r_{out}\geq r_{in}$) and the quintessence horizon $r_q$ ($r_q\geq r_{out}$). For a critical value of the normalization factor $c$, given by
 
\begin{equation}\label{eq.ce}
c_{crit}\epsilon=\frac{1}{6}\sqrt{30\sqrt{5}-66}\approx0.173368\,,
\end{equation} 

the Hayward BH-$\omega$ has one horizon, i.e. $r_{in}=r_{out}=r_q$. Such black holes are called ultra cold black holes \cite{Fernando:2014wma}\cite{Fernando:2013uza}. While, that for $c>c_{crit}$ there is no black hole for any value of the mass $M_{\omega}$, on the other hand, for $c<c_{crit}$ the black hole could have three or two horizons, only for values of the mass $M_{\omega}$ between $M_{\omega min}$ and $M_{\omega max}$, where $M_{\omega min}$ and $M_{\omega max}$ are minimum and maximum local respectively of the expression $M_{\omega}(r)$.

When $M_{\omega}=M_{\omega max}$, we have the case corresponding to the Nariai BH (see \cite{10026018884}). 

\begin{equation}
r_{out}=r_q=-\frac{3c^2\epsilon^2-1}{6c}+\frac{1+30c^2\epsilon^2+9c^4\epsilon^4}{6c\Delta}+\frac{\Delta}{6c}\,,
\end{equation}
where $\Delta$ is given by
\begin{equation}
\Delta=\left(1-117c^2\epsilon^2-135c^4\epsilon^4
-27c^6\epsilon^6+18\sqrt{33c^4\epsilon^4
+9c^6\epsilon^6-c^2\epsilon^2}\right)^{\frac{1}{3}}\,.
\end{equation}

When $M_{\omega}=M_{\omega min}$, we obtain
\begin{equation}
r_{in}=r_{out}=-\frac{3c^2\epsilon^2-1}{6c}-\frac{1}{12c}\left[\frac{1+30c^2\epsilon^2+9c^4\epsilon^4}{\Delta}+\Delta\right]+
i\frac{\sqrt{3}}{12c}\left[\frac{1+30c^2\epsilon^2+9c^4\epsilon^4}{\Delta}-\Delta\right]\,.
\end{equation}
The third term in this expression is purely real for $c<c_{crit}$, and this case corresponding to the cold black hole. When the mass $M_{\omega}$ is between the range $M_{\omega min}<M_{\omega}<M_{\omega max}$, we solve (\ref{ec.rfc}) numerically to find the horizon radius. 
The Fig (\ref{Fig2}). shows both cases: Nariai and cold black holes.  

\section{Geodesics equation}

In this section we will derive null geodesics for the  Hayward BH-$\omega$. Considering the spherical coordinates $x^{\mu}=(t,r,\theta,\phi)$, the test particles propagate along of geodesics are described by the Lagrangian density ${\cal L}= \frac{1}{2}\dot{x}^{\mu}\dot{x}_{\mu}$, where "dot" denotes the derivative with respect to the affine parameter $\tau$. Is possible consider $2{\cal L}=h$ then if $h=1$ correspond to time-like geodesics and $h=0$ correspond to null geodesics.
Then using the form of the line element (\ref{mfa}) and the  expression (\ref{ec.rfc}) the Lagrangian density is given by:

\begin{equation}\label{ec.la}
{\cal L}=\frac{1}{2}\left(
-f_{\omega}(r)\dot t^2+\frac{\dot r^2}{f_{\omega}(r)}+r^2\dot\theta^2+r^2\sin^2\theta\dot\phi^2
\right)\,,
\end{equation}
The motion equation is
\begin{equation}
\dot\Pi_{x^{\mu}}-\frac{\partial{\cal L}}{\partial x^{\mu}}=0\,,
\end{equation}
here $\Pi_{x^{\mu}}=\partial{\cal L}/\partial \dot x^{\mu}$ is the momentum to coordinate $x^{\mu}$. Since the Lagrangian is independent of $t$ and $\phi$, there are two conserved quantities;
\begin{eqnarray}
\Pi_t&=&-f_{\omega}(r)\dot t=-E\label{ecmt}\,,\\
\Pi_{\phi}&=&r^2\sin^2\theta\dot\phi=L\label{ecmp}\,,
\end{eqnarray} 
where $E$ and $L$ are motion constants that corresponding to energy and angular momentum respectively.  

We consider the motion on the plane $\theta=\pi/2$ and using equations (\ref{ecmt}) and (\ref{ecmp}), the Lagrangian in Eq. (\ref{ec.la}) can be written as

\begin{equation}
2{\cal L}\equiv h=\frac{E^2}{f_{\omega}(r)}-\frac{\dot r^2}{f_{\omega}(r)}-\frac{L^2}{r^2}\,.
\end{equation}

Solving for $\dot r^2$, we obtain
\begin{equation}
\dot r^2=E^2-f_{\omega}(r)\left(h+\frac{L^2}{r^2}\right)\,.
\end{equation}

In this paper, we will only focus on null geodesics ($h=0$). Therefore, the equation for the null geodesics for Hayward BH-$\omega$ is given by;

\begin{equation}\label{ec.prpe}
\dot r^2=E^2-V_{\text{eff}}\,,
\end{equation}
or 
\begin{equation}\label{ec.ngu0}
\frac{dr}{d\phi}=\frac{r^2}{L}\sqrt{E^2-V_{\text{eff}}}
\end{equation}

where $V_{\text{eff}}=f_{\omega}(r)\frac{L^2}{r^2}$. The effective potential $V_{\text{eff}}$ of Hayward BH-$\omega$ , can be expanded as:

\begin{equation}
V_{\text{eff}}=
\frac{L^2}{r^2}\left(1-\frac{2M_{\omega}r^2}{r^3+2M_{\omega}\epsilon^2}-cr\right).
\end{equation}

\begin{figure}[h]
\begin{center}
\includegraphics [width =0.7 \textwidth ]{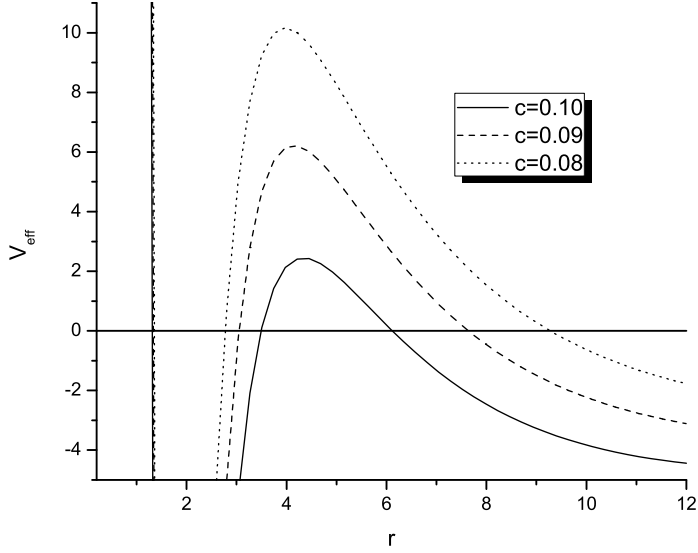}
\end{center}
\caption{The figure shows $V_{eff}$ with $L=20$, $M_{\omega}=1.2$  and $\epsilon=1$} \label{Fig3}
\end{figure}

It is easy to see that $V_{\text{eff}} \rightarrow 0$ when $r \rightarrow \infty$. In the figure (\ref{Fig3}) shows the effective potential of null geodesics for different values of $c$. The shape of the potential is the same, independent of values of $c$. 

Then it is possible to mention that when the parameter $c$ increases to Hayward BH-$\omega$, the effect of the gravitational potential decreases. It is possible to observe the Figure (\ref{Fig3}) maximums of potential then exist unstable null geodesics with a radius of $r_{C}$. Then we can consider three different scenarios depending on the values of $E$ for the motion:

\begin{enumerate}
\item
The first case is $E^2-V_{eff}=0$ when $\dot r=0$, we obtained circular null geodesics. For the null geodesics be  considered as circular orbits stable or unstable,  they must fulfill the following conditions; $V_{eff}^{'}(r_{C}) = V_{eff}(r_{C})=0$ where $r_{C}$ is the radius of the circular orbit and the prime denotes derivative respect to $r$, also for circular stable orbits  the condition $V_{eff}^{''}(r_{C})>0$ must be fulfilled and $V_{eff}^{''}(r_{C})<0$ for unstable. In general the condition for obtain the stables or unstable geodesics is;

\begin{equation}\label{E/L}
\frac{E}{L}=\pm \left[ \frac{1}{2r_{C}}\left(-c+ \frac{2M_{\omega}r_{C}(r_{C}^{3}-4M_{\omega}\epsilon^{2})}{(r_{C}^{3}+2M_{\omega}\epsilon^{2})^{2}}\right)\right]^{1/2}
\end{equation}

for example is possible consider $E=E_{C}$ corresponding to the maximum value of $V_{eff}$ located in a $r_{C}$.

\item
The second case $E_{1}^{2}-V_{eff}>0$ for all $r$. This case denotes the value of the energy $E=E_{1}$ that corresponding to the open null geodesics.

\item
The third case $E_{2}^{2}-V_{eff}<0$ for all $r$. This case denotes the value of the energy $E=E_{2}$ that corresponding to the closes null geodesics.
\end{enumerate}

Also is possible mention that the effective potential, $V_{eff}\to0$ if $r$ tends to $r_{in}$, $r_{out}$ and $r_q$. When two horizons coincide together, the maximum of effective potential will be zero. In this paper, we will only study the behavior of photons for non-degenerate horizons. 

\section{Null geodesics}

In order to analyze the geodesic equation of motion (\ref{ec.ngu0}), we would like to make change of variable, $u=\frac{1}{r}$ to study the orbits. Thus, we can rewrite the Eq. (\ref{ec.ngu0}) as

\begin{equation}\label{ec.ngu}
\left(\frac{du}{d\phi}\right)^2=g(u)\,,
\end{equation}

where $g(u)$ corresponds to
\begin{equation}\label{ec.fu}
g(u)=\frac{E^2}{L^2}-u^2+\frac{2M_{\omega}u^3}{1+2M_{\omega}\epsilon^2u^3}+cu.
\end{equation} 

or
\begin{equation}
g(u)=\frac{-2L^{2}M_{\omega}\epsilon^{2}u^{5}+2L^{2}M_{\omega}c\epsilon^{2}u^{4}+2M_{\omega}\left(E^{2}\epsilon^{2}
+L^{2}\right)u^{3}-L^{2}u^{2}+cL^{2}u+E^{2}
}{L^{2}\left(1+2M_{\omega}\epsilon^{2}u^{3}\right)}
\end{equation}

$g(u)$ can be written as;
\begin{equation}
g(u)=\frac{-2L^{2}M_{\omega}\epsilon^{2}(u-u_{1})(u-u_{2})(u-u_{3})(u-u_{4})(u-u_{5})}{L^{2}\left(1+2M_{\omega}\epsilon^{2}u^{3}\right)}
\end{equation}

Where $u_{i}$ are the roots of $g(u)$ and $u_{1}u_{2}u_{3}u_{4}u_{5}=\frac{E^{2}}{2M_{\omega}L^{2}\epsilon^{2}}$,the geometry of the null geodesics will depend on the roots of the function $g(u)$. Note that for any values of the parameters $M_{\omega}$, $Q$, $\epsilon$, $c$, $E$ and $L$, the function $g(u)\to\pm\infty $ for $u\to\pm\infty$. Also when $u\to 0$, $g(u)\to\frac{E^2}{L^2}$. Therefore, $g(u)$ has three positive roots and two negative roots. 
The geodesics' geometry will depend on the roots of the function $g(u)$. Solving Eq. (\ref{ec.ngu}) numerically with appropriate boundary conditions, one can obtain the geodesic followed by massless particles in the Hayward BH-$\omega$.  

\begin{figure}[h]
\begin{center}
\includegraphics [width =0.47 \textwidth ]{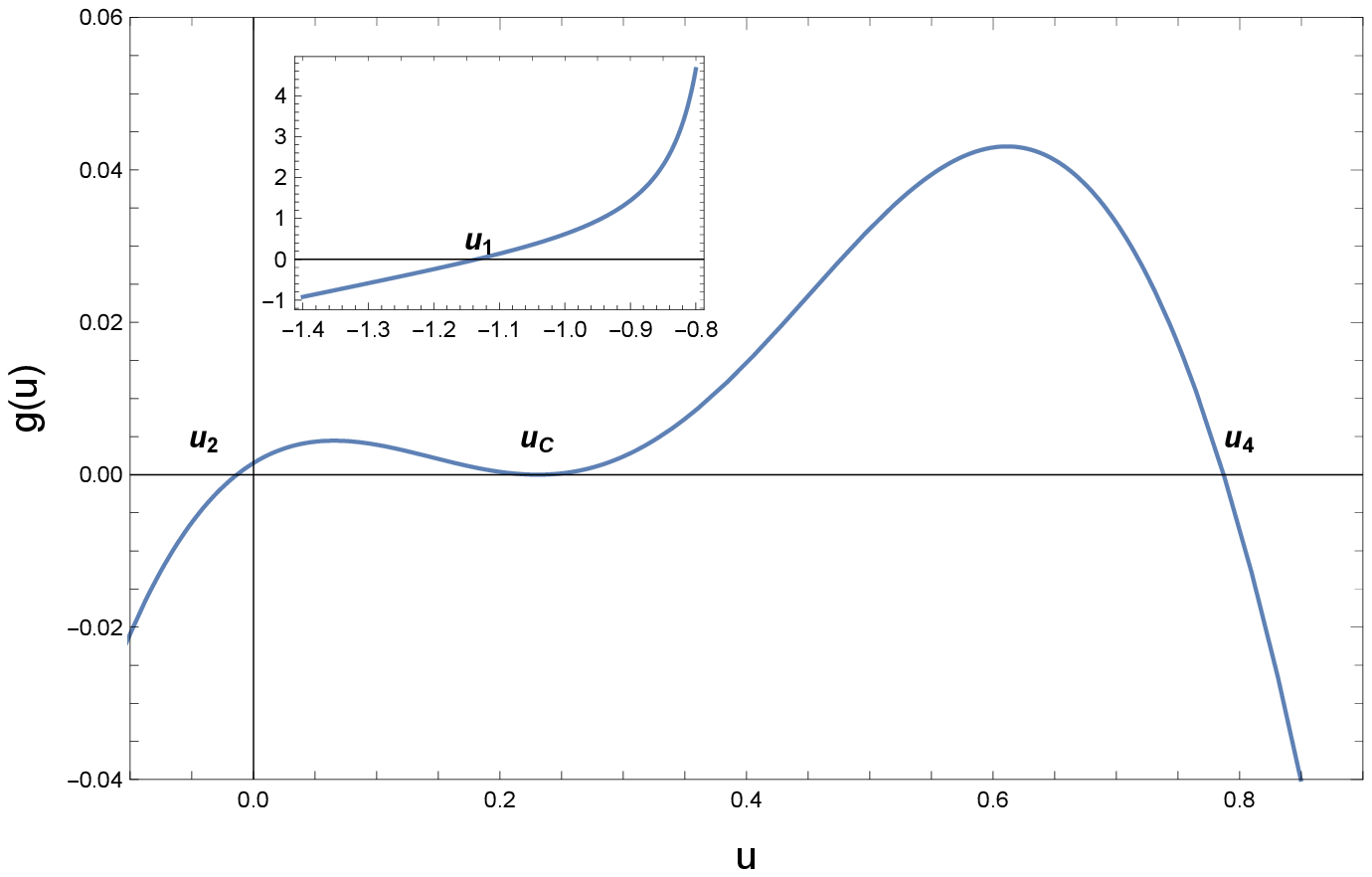}
\includegraphics [width =0.47 \textwidth ]{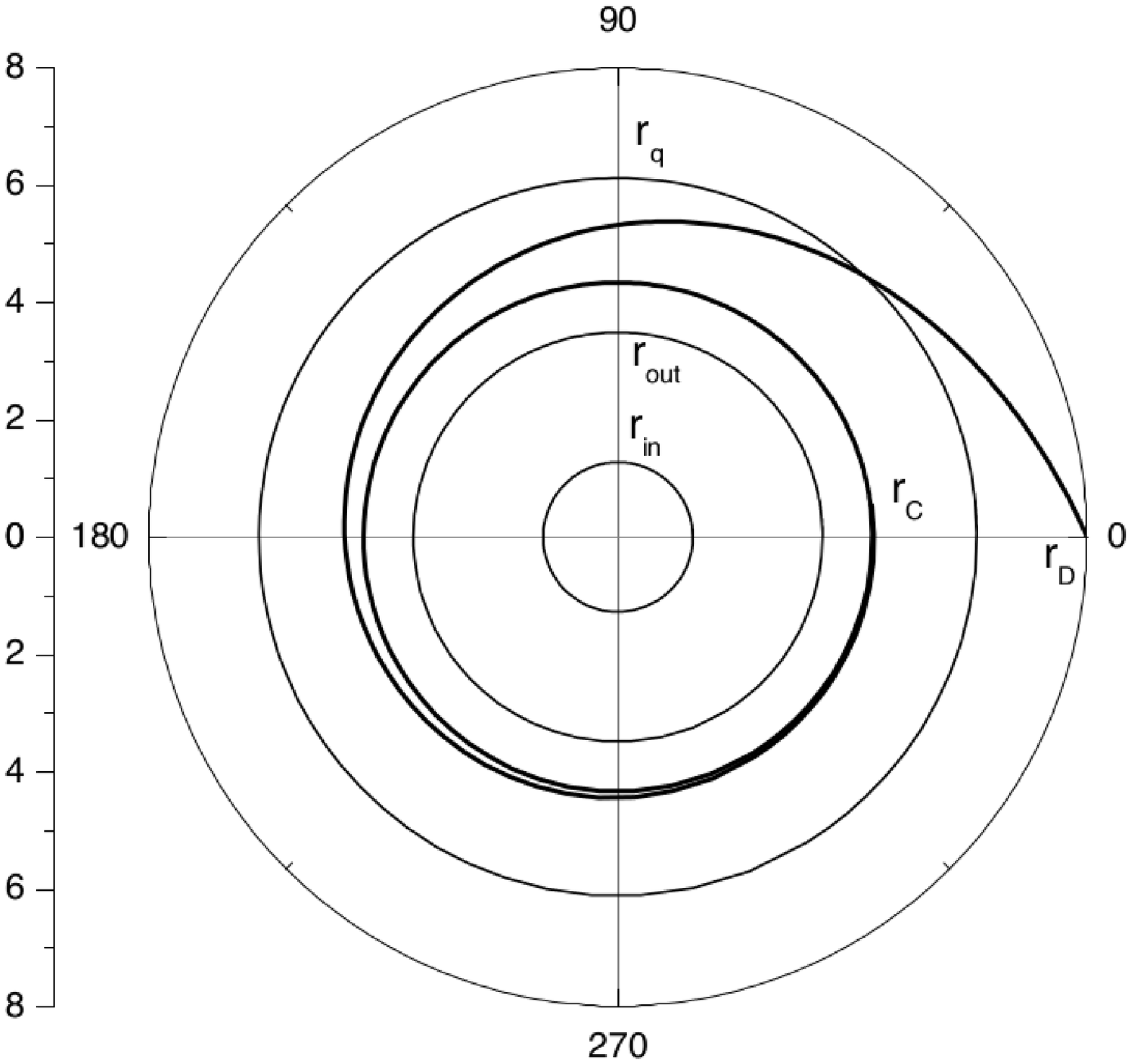}
\end{center}
\caption{The geodesic have an unstable circular orbit at $r=r_C$. Here $E_C=1.00125$, $L=25.6$, $M_{\omega}=1.2$, $\epsilon=1$, $c=0.1$ and $r_C=4.3385$.} \label{Fig4}
\end{figure}

For a test particle (photon) arriving from $r_{D} > r_{C}$ undergo unstable circular orbit at $r=r_{C}$. The corresponding motion is given in the  figure (\ref{Fig4}). When we consider the valuer of $E=E_{C}$ the function $g(u)$ has a degenerate root $u_{C}=\frac{1}{r_{C}}$ as can also be seen in figure (\ref{Fig4}), the graph for $g(u)$ shows four roots.

\begin{figure}[h]
\begin{center}
\includegraphics [width =0.47 \textwidth ]{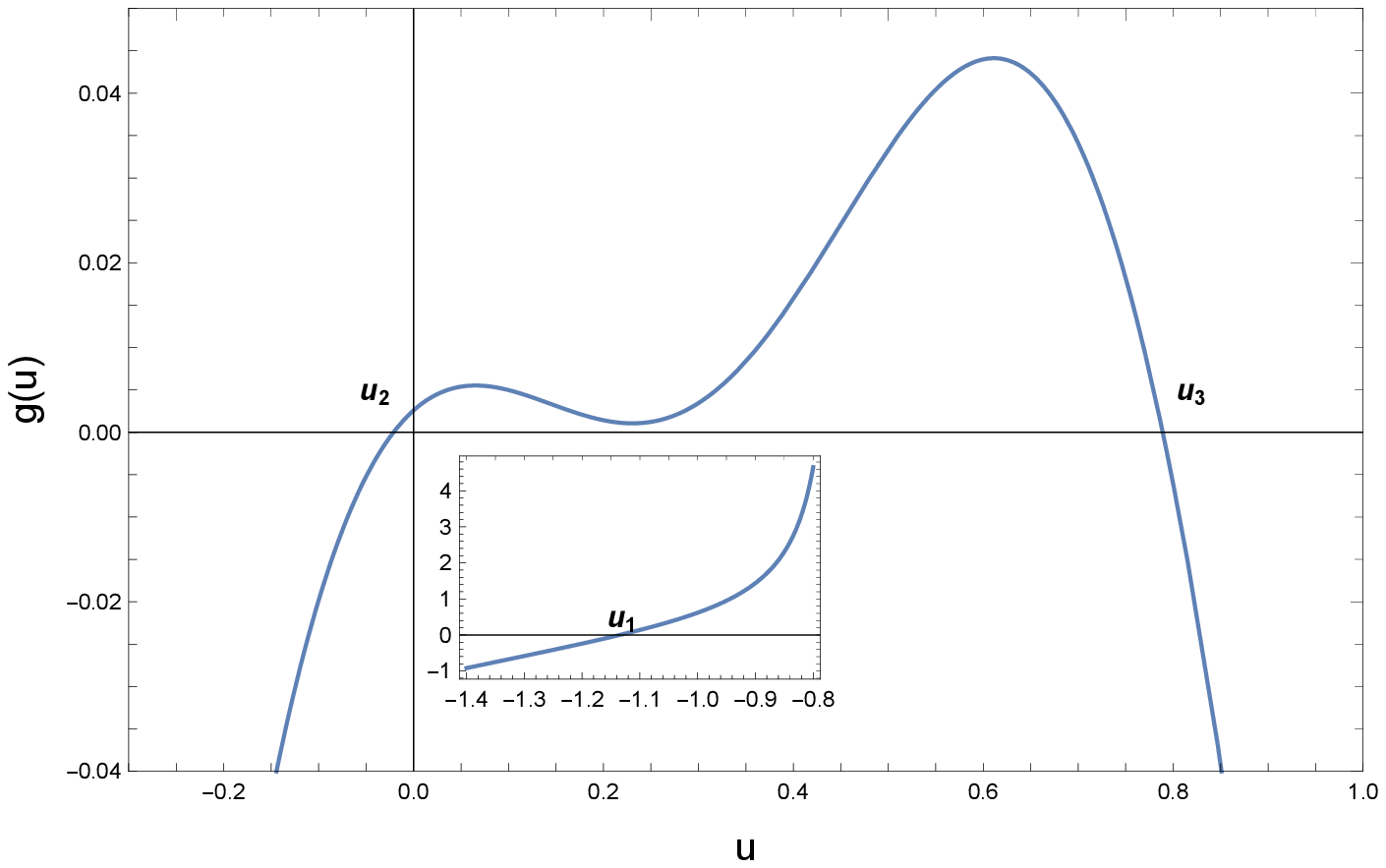}
\includegraphics [width =0.47 \textwidth ]{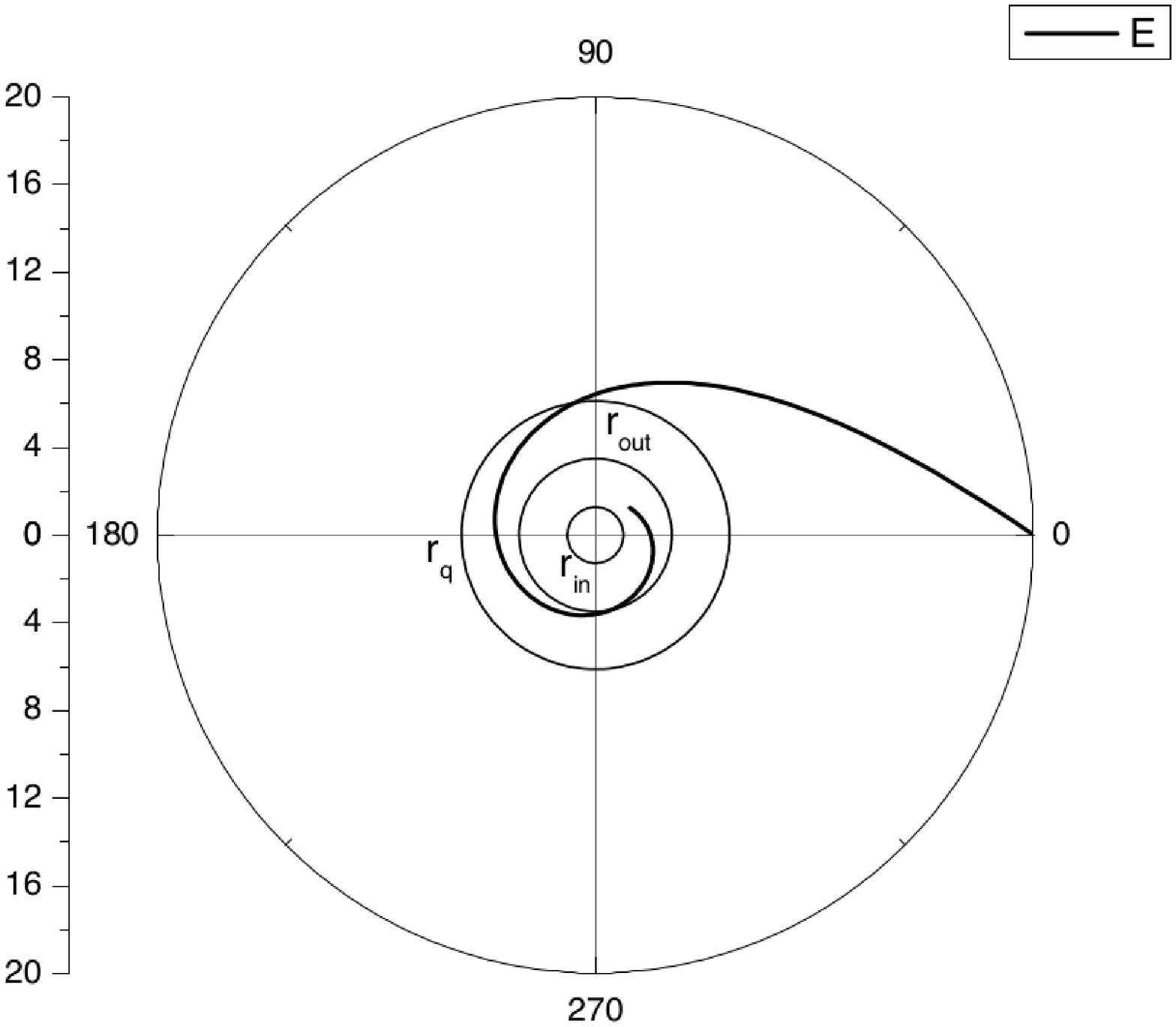}
\end{center}
\caption{The geodesic have an unstable circular orbit at. Here $E_1=1.30125$, $L=25.6$, $M_{\omega}=1.2$, $\epsilon=1$ and $c=0.1$.} \label{Fig5}
\end{figure}

When we consider $E=E_{1}$ for all values of $r$, the photons will fall into the black hole in the figure (\ref{Fig5}) shows this situation. In this situation the function $g(u)$ has two imaginary and three real roots as shown in the graph of $g(u)$ in the figure(\ref{Fig5}).

\begin{figure}[h]
\begin{center}
\includegraphics [width =0.47 \textwidth ]{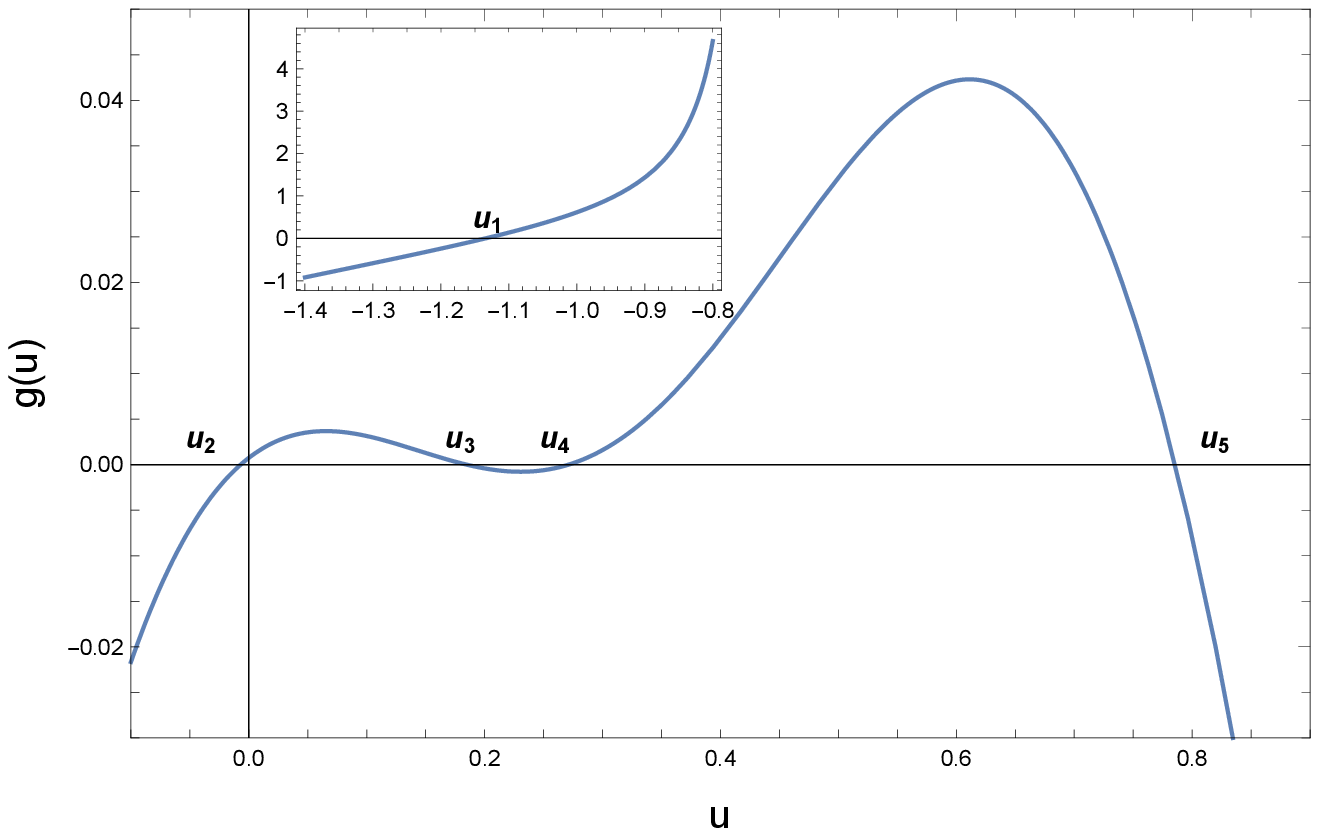}
\includegraphics [width =0.47 \textwidth ]{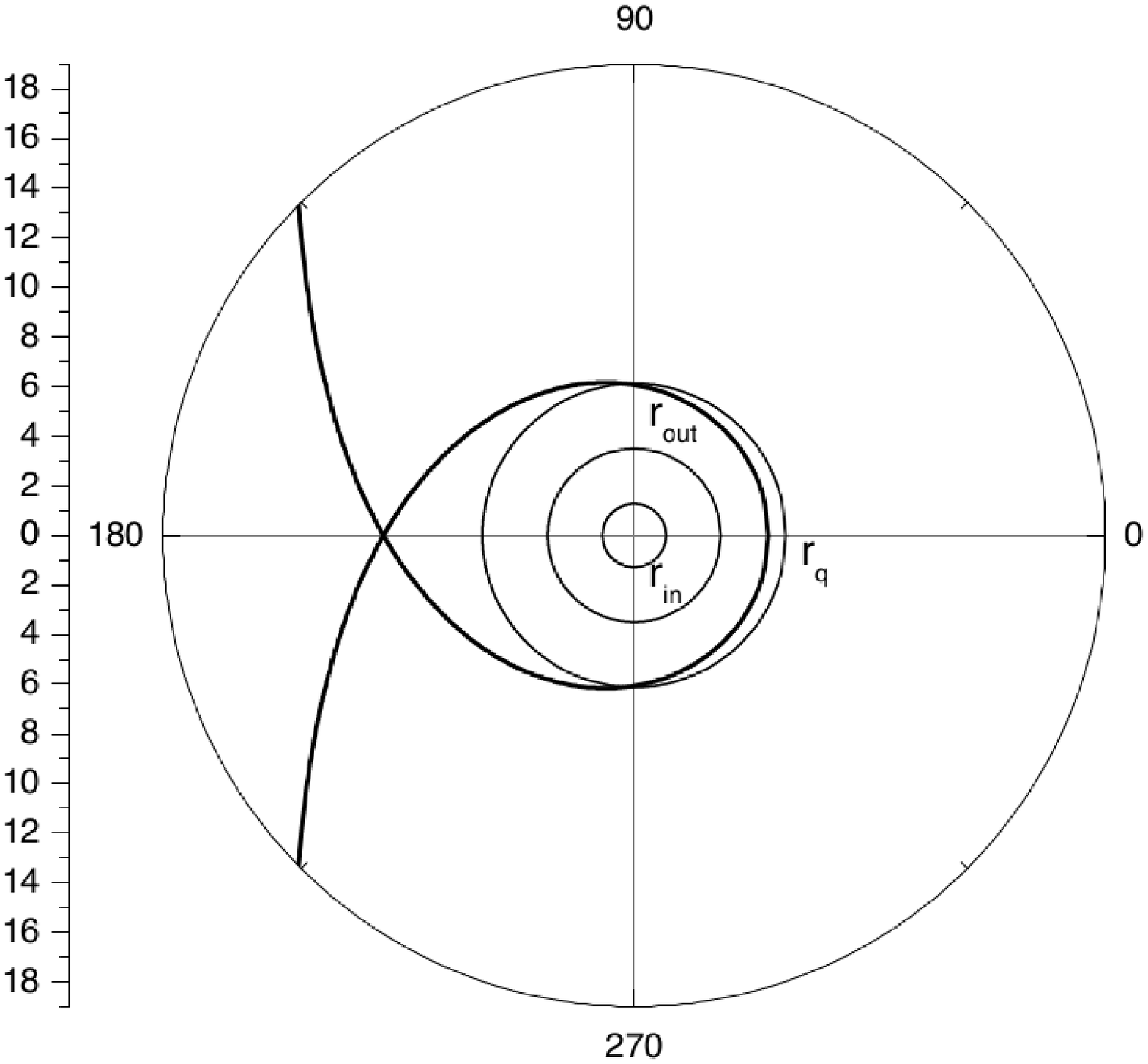}
\end{center}
\caption{The geodesic have an unstable circular orbit at. Here $E_2=0.70125$, $L=25.6$, $M_{\omega}=1.2$, $\epsilon=1$ and $c=0.1$.} \label{Fig6}
\end{figure}

Now when $E=E_{2}$ the test particle starts far form the  Hayward BH-$\omega$ at $u_{D}=1/r_{D}$, it will fall until $u_{out}=1/r_{out}$ and fly away from the black holes, the figure (\ref{Fig6}) shows this situation where $g(u)$ has five real root. 

\section{The shadow of Hayward BH-$\omega$}

The shadow of a black hole in the region on the observer's sky is left dark if the light sources are anywhere in the universe but not between the observer and the black hole. The shadow shape will give important information on the parameters of the black hole.

The behavior of the effective potentials are shown in the figure \ref{Fig3}, then  we observe the existence of photo spheres, since the Hayward BH-$\omega$ is spherically symmetric the shadow will be circularly symmetric and will be a function only of the impact parameter defined as $b_{c}^{2}= L^{2}/E^{2}$ \cite{Shaikh:2018lcc} also the impact parameter $b_{c}$ is related to the effective potential by $V_{ef}(r_{c})=1/b_{c}^{2}$. The relation between the shadow area and impact parameter is $\sigma=\pi b_{c}^{2}$.

From combining the equation (\ref{E/L}) and the definition of shadow area the impact parameter is given by;

\begin{equation}\label{Shad}
\frac{1}{b_{C}^{2}}= \frac{1}{2r_{C}}\left(-c+ \frac{2M_{\omega}r_{C}(r_{C}^{3}-4M_{\omega}\epsilon^{2})}{(r_{C}^{3}+2M_{\omega}\epsilon^{2})^{2}}\right)
\end{equation}

In Fig (2), the behavior of the shadow area of Hayward BH-$\omega$ is shown, as well as the case of Hayward black hole ($c=0$).  Then it is possible to mention that when is introducing the normalization factor $c$ and the quintessence state parameter $\omega$, the shadow area increases ($\sigma_{Hayward BH-\omega}> \sigma_{Hayward BH}$).

\begin{figure}[h]
\begin{center}
\includegraphics [width =0.6 \textwidth ]{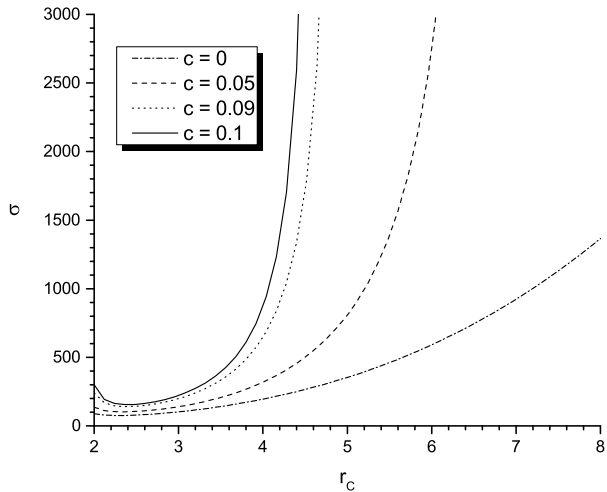}
\end{center}
\caption{The figure shows the shadow area of Hayward BH-$\omega$ with  $M_{\omega}=1.2$  and $\epsilon=1$ for different values of $c$} \label{Fig7}
\end{figure}

So it is possible to conclude that the shadow area increases when the factor $c$ increases, however, the range of $r_{C}$  decrease i.e., exist greater combinations of the parameters $E$ $L$, $M_{\omega}$ and $\epsilon$ for we obtain of photospheres with radius $r_{C}$. In \cite{Abdujabbarov:2016hnw} the study of the shadow of  rotating Hayward black hole is realized and in \cite{doi:10.1139/cjp-2019-0572} mention is made of the shadow of Hayward with charge

\section{Conclusions}

In this paper,  using the method described by Kiselev, we study the metric of regular Hayward black hole surrounded by quintessence at the special case of $\omega=-\frac{2}{3}$. It is possible to observe that an additional horizon (quintessence horizon) can be obtained to the horizons that the black hole of Hayward contains, this new horizon is related to the parameter $\omega$.

A study is made of the horizons of the Hayward black hole surrounded by quintessence. Depending on a critical value of the normalization factor, the Hayward black hole surrounded by quintessence has one, two, or three horizons.

By analyzing the effective potential of test particles (photon),  we have investigated the null geodesics and the kinds of orbits of the Hayward black hole surrounded by quintessence corresponding to different energy levels. The movement of the photons can be located within the quintessence horizon, but if it passes within the horizon $r = r_ {out}$, the photons fall into the black hole, then quintessence horizon is an apparent horizon. Finally is possible to mention that the shadow area increases when the factor $c$ increases.

\section*{ACKNOWLEDGEMENT}

The authors acknowledge the financial support from SNI-CONACYT, México.

\bibliographystyle{unsrt}

\bibliography{bibliografia}

\end{document}